# ASYMPTOTIC DECOMPOSITION IN THE PROBLEM OF JOINED ELASTIC PLATES

**A.G. Kolpakov**
Siberian State University of Telecommunication and Informatics, Kirov str., 86,
Novosibirsk, RU-630009, Russia
E-mail: algk@ngs.ru

**I.V. Andrianov**
**B. Markert**
Institute of General Mechanics, RWTH Aachen University, Templergraben 64,
D-52056, Aachen, Germany
E-mail: andrianov@iam.rwth-aachen.de; markert@iam.rwth-aachen.de

In this paper, a method of local perturbations, previously successfully applied to decompose the problem of elasticity in the system of connected thin rods and beams [Kolpakov and Andrianov, 2013], is used to study the asymptotic behavior of the elasticity problem in connected thin plates. A complete decomposition of the problem, i.e. the separation of the original problem in to the two-dimensional problem of the theory of plates and local problems is proposed. The local problems describe the three-dimensional stress-strain state in the connected plates and can be solved by numerical methods.

## Introduction

In this paper the asymptotic analysis of the three-dimensional problem of the theory of elasticity for two connected thin areas-plates is proposed. The plate and the joint are treated as three-dimensional bodies. The problem is similar to the problem formulation of the deformation of connected rods and beams-frames and trusses in structural mechanics. However, the problems are not entirely identical.

As for the plates and for frames and trusses, the connector assembly may have a very complicated structure. However, for frames and trusses the idealization of the connector assembly is a point for which the equilibrium conditions are written in almost obvious way [Chen and Luis, 2005]. For plates, a joint has the form of a line along which the interaction of the plates (in the simplest case considered here two, and in the general case several plates) takes place. The increasing complexity of problems in the transition to the consideration of the joined plates is similar to the complication of the problem in the transition from the beam theory to the theory of plates that is associated with an increased dimension of the problem.

The difference is due to historical reasons as frames and trusses have been studied on a fairly rigorous mathematical basis for more than two centuries (see [Timoshenko, 1983]).In contrast, plates are for structural mechanics a relatively new subject and now the theory of plates is rather a part of the theory of elasticity than of structural mechanics (at least, the authors do not know the analogue of frame and truss theories with respect to plate structures). Perhaps, the lack of a theory along with increasing practical use of plate structures (plates not only as structural members, but also the use of cellular, foam and other highly porous materials as structural elements, see for example [Banhart et. al., 2001] and the use of plates from composite material [Kalamkarov and Kolpakov, 1996; Andrianov et al., 2004; Kienzler et al., 2004]) resulted in a recent interest in the theoretical analysis of joints between plates (see [Bernadou, 1989, 1995; Ciarlet, 1991; Le Dret, 1991; Titeux and Sanchez-Palencia, 2000]), which does not fall far [http://www. convegni.unicas.it, 2013]. Note, that the problem of joint arises in many other fields of engineering and natural sciences and in mathematics, see [Adler and Mityushev, 2010; Panasenko, 2005; Pokorny et al. 2004] and references in the mentioned publications.

Note that the number of publications devoted to the analysis of connected plates highly exceeds the number of publications devoted to the analysis of connected rods and beams, although the problems arising in the calculation of the local stress-strain state in the connected trusses, according to the authors' opinion, are even more difficult and time-consuming than the analysis of the connected plates. It may not be obvious from the point of view of structural mechanics, but the asymptotic theory indicates that the analysis of the beam leads to the need to address the increasing number of different types of "local" problems, see [Annin et al., 1993; Kalamkarov and Kolpakov, 1997].

The construction of a method for decomposition of an original problem into a "global" (the problem of the theory of plates) and a "local" one (three-dimensional problem of the theory of elasticity) is actual. It is also important to calculate the local stress-strain state at any plate point (first of all in the vicinity of the joint). A similar problem arises in the method of homogenization based on the multi-scale approach [Sanchez-Palencia, 1980; Jikov et al., 1994; Bensoussan et al., 1978], where one proceeds from a split of the original problem in the homogenized problem and the problem in the periodic cell.

For connected plates there are developed asymptotic methods [Ciarlet, 1991; Le Dret, 1991], for which the use of a priori hypotheses about the form of the stress-strain state in the joint vicinity is an important limitation. Methods based on the boundary layer theory [Maz'ya et al., 1981; Motygin and Nazarov, 2000; Gol'denveizer and Kaplunov, 1988], which are effective for describing the contacting plates, cannot be applied in their current form if there is an inset between plates (it is not clear how these methods can be used, for example, for a calculation of the stress-strain state in a bolted connection).

The most significant advance in the study of the problem of connected thin-walled elastic elements was done in works devoted to the method of partial homogenization, see [Panasenko, 2005, 2007], which can be characterized as a partial decomposition of the problem, i.e. a partition of the original problem in to two-dimensional elements (in the main part of the plate) matched with three-dimensional elements for the connector assembly.

In this paper, we propose a method for the complete decomposition of the three-dimensional problem of elasticity in thin connected domains. The method is based on the use of a multi-scale decomposition, effectively used in and known from the theory of homogenization [Sanchez-Palencia, 1980; Jikov et al., 1994; Bensoussan et al., 1978] and the method of local corrector [Gaudiello and Kolpakov, 2011; Kolpakov, 2011]. This raises the problem of the theory of elastic bodies with local perturbations of geometry and material characteristics, similar to a cell problem of the homogenization theory, but where the condition that the "corrector is a periodic function" is replaced by the condition that the "corrector is localized in the vicinity of the perturbation". This difference is significant. In particular, it leads to the individual properties of joints that are not accounted absolutely in the "global" problem corresponding to the engineering theory of structures.

At the same time, in the proposed method, there are still many advantages of the homogenization approach, first of all the possibility of complete decomposition of the problem (the subdivision of the original problem into a two-dimensional "global" and the three-dimensional "local" problem) and the ability to calculate the local stress-strain state in the joint and its vicinity in the form of a simple combination of the solutions of the "global" and the "local" problem.

The obtained local problems can be solved using standard FEM codes. It makes the method of local parameters an effective method for engineering calculations of the stress-strain state in the connector assembly of thin-walled structures.

## 1 Statement of the problem

Consider a three-dimensional domain $\Omega_\varepsilon$, consisting of two rectangular parallelepipeds of a thickness substantially less than the other two dimensions (the plates) and a domain of arbitrary shape connecting the plates (Fig.1). Connection sizes are small in two directions. The main types of joints for thin plates are welded, riveted or bolted, for more details see [Blake, 1985; Collins et al, 2002].

In most engineering structures and components of mechanical devices, the size of a typical joint in the two dimensions is equal to the thickness of the plates. Consider the case where the characteristic thickness of the plate and joint sizes in two directions are proportional to one small parameter $\varepsilon$ (Fig. 1).

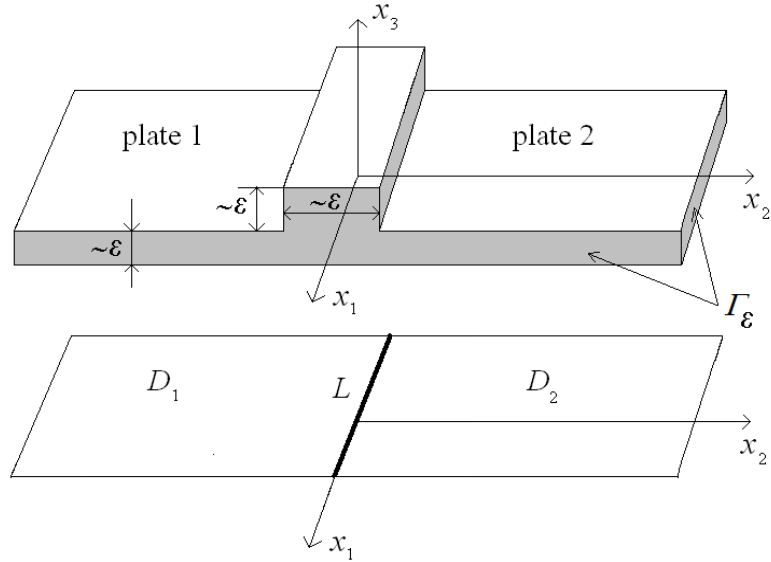

**Fig. 1.** Joined plates and two-dimensional model (line $L$ corresponds to the joint).

We use the following notation: $\mathbf{x} = (x_1, x_2, x_3)$, $\mathbf{X} = (x_1, x_2)$ are coordinates in the plane of the plate. We suppose that the axis $x_3$ is perpendicular to the plane of the plate, see Fig. 1. Latin indices take the values 1, 2, 3; Greek indices take the values 1, 2; repeated indices are summed up (except where otherwise stated).

The problem of the theory of elasticity can be written as a problem of minimizing the functional of additional energy [Timoshenko and Goodier, 1951]

$$L(\mathbf{u}) = \int_{\Omega_\varepsilon} E(\mathbf{x}, \mathbf{u}(\mathbf{x})) d\mathbf{x} + \int_{\Omega_\varepsilon} \mathbf{f}(\mathbf{X})\mathbf{u}(\mathbf{x}) d\mathbf{x} \to \min, \mathbf{u} \in V_\varepsilon, \tag{1}$$

where

$$E(\mathbf{x},\mathbf{u}(\mathbf{x})) = \frac{1}{2}\sum_{k,l=1}^{3}\sigma_{kl}u_{k,l} = \frac{1}{2}\sum_{i,j=1}^{3}a_{ijkl}u_{k,l}u_{i,j}.$$

is the local density of elastic energy; $a_{ijkl}$ – elastic constants;

$$V_\varepsilon = \{\mathbf{u}(\mathbf{x}) \in H^1(\Omega_\varepsilon) : \mathbf{u}(\mathbf{x}) = 0 \quad \text{on } \Gamma_\varepsilon\}$$

is the set of permissible displacements (we assume that the plate is under the influence of body mass forces $\mathbf{f}(\mathbf{X})$ and fixed at the surface $\Gamma_\varepsilon$, where as the other surfaces are stress free).

The method of local perturbations is based on searching a solution to problem (1) as the sum of solutions of homogeneous structural elements (in this case for plates) and the perturbation function. In this case, the solution is sought in the form

$$\mathbf{u}^\varepsilon(\mathbf{x}) = u_\alpha(\mathbf{X})\mathbf{e}_\alpha + w(\mathbf{X})\mathbf{e}_3 - w_{,\alpha x}(\mathbf{X})x_3\mathbf{e}_\alpha + \varepsilon\mathbf{V}(x_3/\varepsilon) + \varepsilon\mathbf{v}(\mathbf{x}/\varepsilon) = \qquad (2)$$
$$= u_\alpha(\mathbf{X})\mathbf{e}_\alpha + w(\mathbf{X})\mathbf{e}_3 - \varepsilon w_{,\alpha x}(\mathbf{X})y_3\mathbf{e}_\alpha + \varepsilon\mathbf{V}(y_3) + \varepsilon\mathbf{v}(\mathbf{y})$$

where the functions $u_\alpha(\mathbf{X})$ ( $\alpha = 1,2$ ) and $w(\mathbf{X})$ depend on the "slow" variable $\mathbf{X} = (x_1, x_2)$ and make sense of the global displacements in the plane of the plate and normal to the plate, respectively; the function $\mathbf{v}(\mathbf{y})$ depends on a "fast" variable $\mathbf{y} = \mathbf{x}/\varepsilon$ and decays rapidly with distance from the geometric perturbation.

The expression $u_\alpha(\mathbf{X})\mathbf{e}_\alpha + w(\mathbf{X})\mathbf{e}_3 - w_{,\alpha x}(\mathbf{X})x_3\mathbf{e}_\alpha$ in (2) describes the displacement of a homogeneous plate in accordance with the classical theory [Timoshenko and Woinowsky-Krieger, 1958]. The term $\varepsilon\mathbf{V}(x_3/\varepsilon)$ is the corrector of the asymptotic theory of plates [Caillerie, 1984; Kohn and Vogelius, 1984] (in the absence of this corrector displacements $u_\alpha(\mathbf{X})\mathbf{e}_\alpha + w(\mathbf{X})\mathbf{e}_3 - w_{,\alpha x}(\mathbf{X})x_3\mathbf{e}_\alpha$ do not satisfy the condition of zero normal stress on the free surface of the plates, i.e. this corrector takes into account the effect of transverse deformation). The term $\varepsilon\mathbf{v}(\mathbf{x}/\varepsilon)$ is the local corrector [Kolpakov and Andrianov, 2013], generated by the joint. Without this local corrector the displacements of the classical theory do not satisfy the condition of zero normal stress on the free surfaces if the free surfaces are not parallel to the $Ox_1x_2$-plane and the equations of equilibrium if the plates and the joint are made of different materials or the joint is assembled from several elements (it is clear that if all of the free surfaces are parallel to the $Ox_1x_2$-plane and the plates and the joint are made of the same material, then we are dealing with a homogeneous plate of constant thickness).

Derivatives of displacements $\mathbf{u}^\varepsilon(\mathbf{x})$ (2) are

$$u_{k,l}^\varepsilon(\mathbf{x}) = u_{\alpha,\beta x}(\mathbf{X})\delta_{k\alpha}\delta_{l\beta} - w_{,\alpha x\beta x}(\mathbf{X})y_3\delta_{k\alpha}\delta_{l\beta} + V_{k,ly}(\mathbf{y}) + v_{k,ly}(\mathbf{y}). \qquad (3)$$

The components of the stress tensor $\sigma_{ij}$, corresponding to the displacements $\mathbf{u}^\varepsilon(\mathbf{x})$ (2) shall be determined, with regard to the formula (3), in the form

$$\sigma_{ij} = a_{ijkl}u_{k,l}^\varepsilon(\mathbf{x}) = a_{ij\alpha\beta}u_{\alpha,\beta x}(\mathbf{X})\delta_{k\alpha}\delta_{l\beta} - a_{ij\alpha\beta}w_{,\alpha x\beta x}(\mathbf{X})x_3\delta_{l\beta}\delta_{k\alpha} + \qquad (4)$$
$$a_{ijkl}(V_{k,ly}(\mathbf{y}) + v_{k,ly}(\mathbf{y})).$$

In (4), the symmetry of the elastic constants is used [Timoshenko and Goodier, 1951].

By using the formula (3), we can write the local energy $E(\mathbf{x}, \mathbf{u}^\varepsilon(\mathbf{x}))$ corresponding to the displacements (2) $\mathbf{u}^\varepsilon(\mathbf{x})$, as

$$E(\mathbf{x}, \mathbf{u}^\varepsilon(\mathbf{x})) = \frac{1}{2} a_{ijkl} u^\varepsilon_{i,j}(\mathbf{x}) u^\varepsilon_{k,l}(\mathbf{x}) = \tag{5}$$

$$\frac{1}{2}\sigma_{kl}\,[u_{\alpha,\beta x}(\mathbf{X})\delta_{k\alpha}\delta_{l\beta} - w_{,\alpha x\beta x}(\mathbf{X})x_3\delta_{k\alpha}\delta_{l\beta} + V_{k,ly}(y_3) + v_{k,ly}(\mathbf{y})] =$$

$$\frac{1}{2}a_{\alpha\beta\gamma\delta}u_{\alpha,\beta x}(\mathbf{X})u_{\gamma,\delta x}(\mathbf{X}) - a_{\alpha\beta\gamma\delta}w_{,\alpha x\beta x}(\mathbf{X})x_3 u_{\gamma,\delta x}(\mathbf{X}) +$$

$$a_{\gamma\delta kl}(V_{k,ly}(y_3) + v_{k,ly}(\mathbf{y}))u_{\gamma,\delta x}(\mathbf{X}) -$$

$$\frac{1}{2}a_{\alpha\beta\gamma\delta}w_{,\alpha x\beta x}(\mathbf{X})w_{,\gamma x\delta x}(\mathbf{X})x_3^2 - a_{ij\alpha\beta}(V_{k,ly}(y_3) + v_{k,ly}(\mathbf{y}))w_{,\alpha x\beta x}(\mathbf{X})x_3 +$$

$$\frac{1}{2}a_{ijkl}(V_{k,ly}(y_3) + v_{k,ly}(\mathbf{y}))(V_{i,jy}(y_3) + v_{i,jy}(\mathbf{y})) = E(u_1, u_2, w, \mathbf{v}, \mathbf{V}).$$

After substituting (5) into (1), the minimization problem (1) takes the form

$$L(u_1, u_2, w, \mathbf{v}, \mathbf{V}) = \int_{\Omega_\varepsilon} E(u_1, u_2, w, \mathbf{v}, \mathbf{V}) d\mathbf{x} + \int_{\Omega_\varepsilon} f_\alpha(\mathbf{X})u_\alpha(\mathbf{X}) + f_3(\mathbf{X})w(\mathbf{X})]d\mathbf{x} \to \min, \tag{6}$$

with respect to the functions $u_1(\mathbf{X})$, $u_2(\mathbf{X}) \in W$, $w \in W_2$, $\mathbf{v}(\mathbf{y}) \in U$, and $\mathbf{V}(\mathbf{y}) \in U_h$, where

$$W_1 = \{u(\mathbf{X}) \in H^1(D) : u(\mathbf{X}) = 0 \text{ on } \partial D\}$$

is the set of admissible displacements in the plane of the plate,

$$W_2 = \{u(\mathbf{X}) \in H^2(D) : u(\mathbf{X}) = \frac{\partial w}{\partial \mathbf{n}} = 0 \text{ on } \partial D\}$$

is the set of admissible deflections,

$$U = \{\mathbf{v}(\mathbf{y}) \in H^1(Q_e) : \mathbf{v}(\mathbf{y}) \text{ is localized near the joint}\}$$

is the set of admissible local displacements in the joint and its vicinity, and $U_h = \{\mathbf{V}(\mathbf{y}) \in H^1(Q_\varepsilon)\}$. The listed functional spaces are directly related to the mechanical essence of the problem and are used below in the integration of variation equations in parts.

In (6), terms of the order $\varepsilon$ are omitted.

The Euler-Lagrange equations for the functional $L(u_1, u_2, w, \mathbf{v}, \mathbf{V})$ (6) have the form

$$\delta L_\mathbf{v} = \int_{\Omega_\varepsilon} \{a_{\gamma\delta kl}u_{\gamma,\delta x}(\mathbf{X}) - a_{ij\alpha\beta}w_{,\alpha x\beta x}(\mathbf{X})y_3 + a_{ijkl}(V_{i,jy}(\mathbf{y}) + v_{i,jy}(\mathbf{y}))]\delta v_{k,ly}(\mathbf{y}) d\mathbf{x} = 0, \tag{7}$$

$$\delta L_\mathbf{V} = \int_{\Omega_\varepsilon} [a_{\gamma\delta kl}u_{\gamma,\delta x}(\mathbf{X}) - a_{ij\alpha\beta}w_{,\alpha x\beta x}(\mathbf{X})y_3 + a_{ijkl}(V_{i,jy}(\mathbf{y}) + v_{i,jy}(\mathbf{y}))]\delta V_{k,ly}(\mathbf{y}) d\mathbf{x} = 0 \tag{8}$$

and

$$\delta L_{u_\gamma} = \int\limits_{\Omega_\varepsilon} [-a_{\gamma\delta\alpha\beta} w_{,\alpha x\beta x}(\mathbf{X})x_3 + a_{\gamma\delta kl}(V_{k,ly}(\mathbf{y})+v_{k,ly}(\mathbf{y}))]\delta u_{\gamma,\delta x}(\mathbf{X})d\mathbf{x} + \qquad (9)$$

$$\int\limits_{\Omega_\varepsilon} f_\gamma(\mathbf{X})\delta u_{\gamma,\delta x}(\mathbf{X})d\mathbf{x} = 0,$$

$$\delta L_w = \int\limits_{\Omega_\varepsilon} [-a_{\alpha\beta\gamma\delta}x_3 u_{\gamma,\delta x}(\mathbf{X}) - a_{\alpha\beta\gamma\delta}w_{,\gamma x\delta x}(\mathbf{X})x_3^2 - \qquad (10)$$

$$a_{ij\alpha\beta}(V_{k,ly}(\mathbf{y})+v_{k,ly}(\mathbf{y}))y_3]\delta w_{,\alpha x\beta x}(\mathbf{X})d\mathbf{x} + \int\limits_{\Omega_\varepsilon} f_3(\mathbf{X})\delta w(\mathbf{X})d\mathbf{x} = 0.$$

Here $\delta L$ means the variation with respect to the corresponding variable (the variable is indicated as a subscript).

Next, let us introduce the function $\mathbf{W}(\mathbf{y}) = \mathbf{v}(\mathbf{y}) + \mathbf{V}(y_3)$. Adding the equations (7) and (8), one obtains the variational equation with respect to $\mathbf{W}(\mathbf{y})$. Integrating this equation by parts, we obtain the following boundary value problem:

$$\begin{cases} (a_{ijkl}W_{k,ly}(\mathbf{y}) + a_{\gamma\delta kl}u_{\gamma,\delta x}(\mathbf{X}) - a_{ij\alpha\beta}w_{,\alpha x\beta x}(\mathbf{X})y_3\}_{,jy} = 0 \text{ in } \Omega_\varepsilon, \\ \{a_{ijkl}W_{k,ly}(\mathbf{y}) + a_{\gamma\delta kl}u_{\gamma,\delta x}(\mathbf{X}) - a_{ij\alpha\beta}w_{,\alpha x\beta x}(\mathbf{X})y_3\}n_j = 0 \text{ on free surface}, \\ \mathbf{W}(\mathbf{x}/\varepsilon) \to \mathbf{V}(y_3) \text{ far from the joint.} \end{cases} \qquad (11)$$

Since in the "main" part of the plates (we call the "main" part of the plate, the part located away from the junction) $\mathbf{v}(\mathbf{y}) = 0$, we obtain from (11) the following problem for the function $\mathbf{V}(\mathbf{y})$

$$\frac{d}{dy_3}\{a_{i3k3}\frac{dV_k}{dy_3}(y_3) + a_{i3\gamma\delta}u_{\gamma,\delta x}(\mathbf{X}) - a_{i3\alpha\beta}w_{,\alpha x\beta x}(\mathbf{X})y_3\} = 0 \text{ at } -\varepsilon/2 < y_3 < \varepsilon/2, \qquad (12)$$

$$a_{i3k3}\frac{dV_k}{dy_3}(y_3) + a_{i3\gamma\delta}u_{\gamma,\delta x}(\mathbf{X}) - a_{i3\alpha\beta}w_{,\alpha x\beta x}(\mathbf{X})y_3 = 0 \text{ at } y_3 = \pm\varepsilon/2$$

(we assume that the plate's thicknesses are equal to $\varepsilon$ and are arranged symmetrically relative to the plane $y_3 = 0$).

We use the idea of the multi-scale method [Sanchez-Palencia, 1980; Jikov et al., 1994; Bensoussan et al., 1978] and separate in (11) the functions of "fast" $\mathbf{y}$ and "slow" $\mathbf{X}$ variables.

Problem (11) is interpreted as a problem in the "fast" variables $\mathbf{y}$, in which the functions of "slow" variables $\mathbf{X}$ are included as parameters. Due to the linearity of problem (11), its solution can be written as follows:

$$\mathbf{W}(\mathbf{x}/\varepsilon) = \sum_{\alpha,\beta=1}^{2}[\varepsilon\mathbf{N}^{(0\alpha\beta)}(\mathbf{x}/\varepsilon)u_{\alpha,\beta x}(\mathbf{X}) - \varepsilon\mathbf{N}^{(1\alpha\beta)}(\mathbf{x}/\varepsilon)w_{,\alpha x\beta x}(\mathbf{X})], \qquad (13)$$

where the function $\mathbf{N}^{(\nu\alpha\beta)}(\mathbf{x}/\varepsilon)$ ($\nu = 0,1, \alpha,\beta = 1,2$) are the problem solutions in the "long" domain of the unit height $\Omega_\infty = \frac{1}{\varepsilon}\Omega_\varepsilon$:

$$\begin{cases} \{a_{ijkl} N_{k,ly}^{(\nu\alpha\beta)}(\mathbf{y}) - a_{ij\alpha\beta} y_3^\nu\}_{,jy} = 0 \text{ in } \Omega_\infty, \\ \{a_{ijkl} N_{k,ly}^{(\nu\alpha\beta)}(\mathbf{y}) - a_{ij\alpha\beta} y_3^\nu\} n_j = 0 \text{ on free surface of domain } \Omega_\infty, \\ \mathbf{N}^{(\nu\alpha\beta)}(\mathbf{y}) \to \mathbf{n}^{(\nu\alpha\beta)}(y_3) \text{ far from the joint.} \end{cases} \quad (14)$$

For the solution of problem (12), we place a representation similar to (13) with the functions $\mathbf{n}^{(\nu\alpha\beta)}(\mathbf{x}/\varepsilon)$ ($\nu = 0,1$, $\alpha = 1,2$, $\beta = 1,2$) determined by solving the problem

$$\begin{cases} \dfrac{d}{dy_3}\{a_{i3k3} \dfrac{dn_k^{(\nu\alpha\beta)}}{dy_3}(y_3) - a_{i3\alpha\beta} y_3^\nu\} = 0 \text{ for } -1/2 < y_3 < 1/2, \\ a_{i3k3} \dfrac{dn_k^{(\nu\alpha\beta)}}{dy_3}(y_3) - a_{i3\alpha\beta} y_3^\nu = 0 \text{ for } y_3 = \pm 1/2. \end{cases} \quad (15)$$

The problem (15) is a boundary value problem for ordinary differential equations which has exact solutions for all ($\nu = 0,1$, $\alpha = 1,2$, $\beta = 1,2$) [Kolpakov, 2004], see also Appendix.

We call problems (14) and (15) "the local problems". Local problems are analogous to cell problems of the homogenization theory [Sanchez-Palencia, 1980; Jikov et al., 1994; Bensoussan et al., 1978]). Problem (15) is known from the asymptotic theory of plates (see, e.g., [Caillerie, 1984; Kohn and Vogelius, 1984]). The problem (14) is new. Although the first two equations in (14) coincide with the equations of cell problems in the theory of homogenization in the version for the plates [Caillerie, 1984; Kohn and Vogelius, 1984], but the condition of the theory of homogenization that the "solution of the cell problem is a periodic function" has been replaced in (14) by the condition "$\mathbf{N}^{(\nu\alpha\beta)}(\mathbf{y}) \to \mathbf{n}^{(\nu\alpha\beta)}(y_3)$ far from the joint". In this connection, the term "cell problem" is replaced with the term "local problem".

Substituting (13) into (9) and (10), one obtains the two-dimensional equations of plate deformation (deformation in the plane of the plates and normal deflection)

$$\int_{\Omega_\varepsilon} (-a_{\gamma\delta\alpha\beta} w_{,\alpha\beta x}(\mathbf{X}) x_3 + a_{\gamma\delta kl} [\mathbf{N}^{(0\alpha\beta)}(\mathbf{x}/\varepsilon) u_{\alpha,\beta x}(\mathbf{X}) - \quad (16)$$

$$N^{(1\alpha\beta)}(\mathbf{x}/\varepsilon) w_{,\alpha\beta x}(\mathbf{X})]_{k,ly}) \delta u_{\gamma,\delta x}(\mathbf{X}) d\mathbf{x} + \int_{\Omega_\varepsilon} f_\gamma(\mathbf{X}) \delta u_{\gamma,\delta x}(\mathbf{X}) d\mathbf{x} = 0,$$

$$\int_{\Omega_\varepsilon} (-a_{\alpha\beta\gamma\delta} x_3 u_{\gamma,\delta x}(\mathbf{X}) - a_{\alpha\beta\gamma\delta} w_{,\gamma x\delta x}(\mathbf{X}) x_3^2 - a_{ij\alpha\beta} [\mathbf{N}_{i,jy}^{(0\alpha\beta)}(\mathbf{x}/\varepsilon) u_{\alpha,\beta x}(\mathbf{X}) - \quad (17)$$

$$a_{ij\alpha\beta} \mathbf{N}^{(1\alpha\beta)}(\mathbf{x}/\varepsilon) w_{,\alpha\beta x}(\mathbf{X}))]_{i,jy} y_3) \delta w_{,\alpha\beta x}(\mathbf{X}) d\mathbf{x} + \int_{\Omega_\varepsilon} f_3(\mathbf{X}) \delta w(\mathbf{X}) d\mathbf{x} = 0.$$

Note that for obtaining non-degenerate equations (9) should be divided by $\varepsilon$ and (10) by $\varepsilon^3$ (without this normalization, we arrive at the correct, but meaningless limit: $0 \to 0$ as $\varepsilon \to 0$). The quantities $\varepsilon$ and $\varepsilon^3$ are the orders of the stiffness characteristic of the plate thickness $\varepsilon$, the in-plane and bending stiffness, respectively [Timoshenko and Woinowsky-Krieger, 1958]. It explains the physical meaning of the normalization.

Equations (16) and (17) with the functions $\mathbf{n}^{(\nu\alpha\beta)}(y_3)$ instead of $\mathbf{N}^{(\nu\alpha\beta)}(\mathbf{y})$ arise in the asymptotic theory of plates with no joints, see, e.g., [Caillerie, 1984; Kohn and Vo-gelius, 1984]. Thus, the non trivial terms in (16) and (17) (which do not arise in the asymptotic theory of plates with no joints) are integrals

$$\frac{1}{\varepsilon}\int_{\Omega_\varepsilon} a_{\gamma\delta kl}(N^{0\alpha\beta}_{k,ly}(\mathbf{y}) - n^{0\alpha\beta}_{k,ly}(\mathbf{y}))\delta u_{\gamma,\delta x}(\mathbf{X})d\mathbf{x}, \tag{18}$$

$$\frac{1}{\varepsilon^3}\int_{\Omega_\varepsilon} a_{ij\alpha\beta}(N^{1\alpha\beta}_{i,jy}(\mathbf{y}) - n^{1\alpha\beta}_{i,jy}(y_3))y_3\delta w_{,\alpha x\beta x}(\mathbf{X})d\mathbf{x}. \tag{19}$$

The integrals (18) and (19) tend to zero as $\varepsilon \to 0$. To prove this, note that the integrands in (18) and (19), which have the form $F_{kl}(\mathbf{y})y_3^\delta$ ($\delta = 0,1$), are integrable in $R^3$ (physically it means that the elastic energy corresponding to the deformations $F_{kl}(\mathbf{y})y_3^\delta$ is finite). Following the change of variables $\mathbf{y} = \mathbf{x}/\varepsilon$, one obtains

$$\int_{\Omega_\varepsilon} F_{kl}(\mathbf{x}/\varepsilon)(x_3/\varepsilon)^\mu d\mathbf{x} = \varepsilon^{3+\mu}\int_{\Omega_\varepsilon} F_{kl}(\mathbf{y})y_3^\mu d\mathbf{y} \leq \varepsilon^{3+\mu}c,$$

where the constant $c < \infty$ is the same for all indices $\mu\alpha\beta$ and $kl$.

For the integral in (18) $\mu = 0$, and the limit value (18) for $\varepsilon \to 0$ equals zero. For the integral (19) $\mu = 1$ and the limit value (19) is also zero. Then, substituting in (9) and (10) the function $\mathbf{W}(\mathbf{x}/\varepsilon)$ (13) and function

$$\mathbf{w}(\mathbf{x}/\varepsilon) = \sum_{\alpha,\beta=1}^{2}[\varepsilon \mathbf{n}^{(0\alpha\beta)}(\mathbf{x}/\varepsilon)u_{\alpha,\beta x}(\mathbf{X}) - \varepsilon \mathbf{n}^{(1\alpha\beta)}(\mathbf{x}/\varepsilon)w_{,\alpha x\beta x}(\mathbf{X})] \tag{20}$$

($\mathbf{n}^{(\nu\alpha\beta)}$ is the solution to (15)), we obtain in the limit (at $\varepsilon \to 0$) the same result, viz.

$$\int_{\Omega_\varepsilon}(-a_{\gamma\delta\alpha\beta}w_{,\alpha x\beta x}(\mathbf{X})x_3 + \tag{21}$$

$$a_{\gamma\delta kl}[\mathbf{n}^{(0\alpha\beta)}(\mathbf{x}/\varepsilon)u_{\alpha,\beta x}(\mathbf{X}) - \mathbf{n}^{(1\alpha\beta)}(\mathbf{x}/\varepsilon)w_{,\alpha x\beta x}(\mathbf{X})]_{k,ly})\delta u_{\gamma,\delta x}(\mathbf{X})d\mathbf{x} +$$

$$\int_{\Omega_\varepsilon} f_\gamma(\mathbf{X})\delta u_{\gamma,\delta x}(\mathbf{X})d\mathbf{x} = 0 \quad (\gamma = 1,2),$$

$$\int_{\Omega_\varepsilon}(-a_{\alpha\beta\gamma\delta}x_3 u_{\gamma,\delta x}(\mathbf{X}) - a_{\alpha\beta\gamma\delta}w_{,\gamma x\delta x}(\mathbf{X})x_3^2 -$$

$$a_{ij\alpha\beta}[\mathbf{n}^{(0\alpha\beta)}(\mathbf{x}/\varepsilon)u_{\alpha,\beta x}(\mathbf{X}) - \mathbf{n}^{(1\alpha\beta)}(\mathbf{x}/\varepsilon)w_{,\alpha x\beta x}(\mathbf{X})]_{k,ly})\delta w_{,\alpha x\beta x}(\mathbf{X})d\mathbf{x} + \tag{22}$$

$$\int_{\Omega_\varepsilon} f_3(\mathbf{X})\delta w(\mathbf{X})d\mathbf{x} = 0.$$

It is known (see, e.g., [Kalamkarov and Kolpakov, 1997]) that from (21) and (22) follow the equations of the classical theory of homogeneous flat plates [Timoshenko and Woinowsky-Krieger, 1958] and the equation of the asymptotic theory of plates [Caillerie, 1984; Kohn and Vogelius, 1984]. Despite the fact that in the classical theory of plates no analogue of the function $\mathbf{V}(y_3)$ exists (as a result, the well-known contradiction arises between the hypothesis of undeformed normal and zero stresses on the free surfaces of the plate), equations of the classical theory of plates coincide with the equations of the asymptotic theory written for homogeneous flat plates (see, e.g., [Kolpakov, 2004]).

$$\int_{\Omega_\varepsilon} [-a_{\gamma\delta\alpha\beta} w_{,\alpha x \beta x}(\mathbf{X}) x_3 +$$

$$a_{\gamma\delta kl}[\mathbf{n}^{(0\alpha\beta)}(\mathbf{x}/\varepsilon) u_{\alpha,\beta x}(\mathbf{X}) - \mathbf{n}^{(1\alpha\beta)}(\mathbf{x}/\varepsilon) w_{,\alpha x \beta x}(\mathbf{X})]_{k,ly} \delta u_{\gamma,\delta x}(\mathbf{X}) d\mathbf{x} + \quad (23)$$

$$\int_{\Omega_\varepsilon} f_\gamma(\mathbf{X}) \delta u_{\gamma,\delta x}(\mathbf{X}) d\mathbf{x} = 0 \quad (\gamma = 1,2),$$

$$\int_{\Omega_\varepsilon} [-a_{\alpha\beta\gamma\delta} x_3 u_{\gamma,\delta x}(\mathbf{X}) - a_{\alpha\beta\gamma\delta} w_{,\gamma x \delta x}(\mathbf{X}) x_3^2 - \quad (24)$$

$$a_{ij\alpha\beta}[\mathbf{n}^{(0\alpha\beta)}(\mathbf{x}/\varepsilon) u_{\alpha,\beta x}(\mathbf{X}) - \mathbf{n}^{(1\alpha\beta)}(\mathbf{x}/\varepsilon) w_{,\alpha x \beta x}(\mathbf{X})]_{k,ly} \delta w_{,\alpha x \beta x}(\mathbf{X}) d\mathbf{x} +$$

$$\int_{\Omega_\varepsilon} f_3(\mathbf{X}) \delta w(\mathbf{X}) d\mathbf{x} = 0.$$

After completing in (21) and (22) the integration over the fast variables, we obtain the standard variational equations of the two-dimensional plate

$$\int_D [-D_{\gamma\delta\alpha\beta}(\mathbf{X}) w_{,\alpha x \beta x}(\mathbf{X}) + C_{\gamma\delta\alpha\beta}(\mathbf{X}) u_{\alpha,\beta x}(\mathbf{X})] \delta u_{\gamma,\delta x}(\mathbf{X}) d\mathbf{X} + \int_D f_\gamma(\mathbf{X}) \delta u_\gamma(\mathbf{X}) d\mathbf{X} = 0, \quad (25)$$

$$\int_D [-C_{\alpha\beta\gamma\delta}(\mathbf{X}) u_{\gamma,\delta x}(\mathbf{X}) + A_{\alpha\beta\gamma\delta}(\mathbf{X}) w_{,\alpha x \beta x}(\mathbf{X})] \delta w_{,\alpha x \beta x}(\mathbf{X}) d\mathbf{X} + \int_D f_3(\mathbf{X}) \delta w(\mathbf{X}) d\mathbf{X} = 0. \quad (26)$$

The integration in (21) and (22) used the fact that the integrals of the product of functions of the variables $\mathbf{X} = (x_1, x_2)$ and functions of the variable $y_3$ are equal to the product of the integrals of the relevant factors, for more details see [Caillerie, 1984; Kohn and Vogelius, 1984; Kalamkarov and Kolpakov, 1997].

In (25) and (26) $A_{\alpha\beta\gamma\delta}(\mathbf{X})$ and $D_{\gamma\delta\alpha\beta}(\mathbf{X})$ are the membrane and flexural stiffnesses, $C_{\alpha\beta\gamma\delta}(\mathbf{X})$ is the non-symmetric stiffness of the plate. These values are constant within each of the connected plates, but in general different for different plates.

Integrating by parts in (25) and (26), we obtain the two-dimensional equilibrium equations in domains $D_1$ and $D_2$ (see notations in Fig.1)

$$[-D_{\gamma\delta\alpha\beta}(\mathbf{X}) w_{,\alpha x \beta x}(\mathbf{X}) + C_{\gamma\delta\alpha\beta}(\mathbf{X}) u_{\alpha,\beta x}(\mathbf{X})]_{,\gamma x \delta x} = f_\gamma(\mathbf{X}) \text{ in } D_i \ (i = 1,2), \quad (27)$$

$$[-C_{\alpha\beta\gamma\delta}(\mathbf{X}) u_{\gamma,\delta x}(\mathbf{X}) + A_{\alpha\beta\gamma\delta}(\mathbf{X}) w_{,\alpha x \beta x}(\mathbf{X})]_{,\beta x} = f_3(\mathbf{X}) \text{ in } D_i \ (i = 1,2) \quad (28)$$

and the interface condition on the line $L$ between the domains: the in-line jumps of the values

$$[-D_{\gamma\delta\alpha\beta}(\mathbf{X}) w_{,\alpha x \beta x}(\mathbf{X}) + C_{\gamma\delta\alpha\beta}(\mathbf{X}) u_{\alpha,\beta x}(\mathbf{X})] n_\delta,$$

$$[-D_{\gamma\delta\alpha\beta}(\mathbf{X}) w_{,\alpha x \beta x}(\mathbf{X}) + C_{\gamma\delta\alpha\beta}(\mathbf{X}) u_{\alpha,\beta x}(\mathbf{X})]_{,\delta x} n_\gamma,$$

$$[-C_{\alpha\beta\gamma\delta}(\mathbf{X}) u_{\gamma,\delta x}(\mathbf{X}) + A_{\alpha\beta\gamma\delta}(\mathbf{X}) w_{,\alpha x \beta x}(\mathbf{X})] n_\delta$$

equal zero (here $(n_1, n_2)$ indicates the vector-normal to the connecting line of plates lying in the plane of the plates). As one can see, the condition on the connecting line does not take into account the individual properties of the joint. The interface condition on the line of the joint includes stiffnesses of the "inner" parts of the plates only.

If we take into account that the first expressions in square brackets are moments $M_{\gamma\delta}(\mathbf{X})$, the second expressions in square brackets are transverse shear forces $Q_{\gamma}(\mathbf{X})$ and the third expressions in square brackets are in-plane forces $N_{\alpha\beta}(\mathbf{X})$ in the plane of the plate, we find that on the contact line we have zero jumps in $M_{\gamma\delta}(\mathbf{X})n_{\delta}$, $Q_{\gamma}(\mathbf{X})n_{\gamma}$, and $N_{\alpha\beta}(\mathbf{X})n_{\delta}$. This condition also does not take into account the individual properties of the joint. If two identical plates are connected, then (23) and (24) yield one equation with constant coefficients in the whole domain, as if "there is no connection". This is the effect of "erased connections" in the limit problem, formulated explicitly in [Gaudiello and Kolpakov, 2011] (we will discuss this in the example in Section 4).

Let us denote $\sigma_{ij}^{(0\alpha\beta)}(\mathbf{x}/\varepsilon) = a_{ijkl} n_{k,ly}^{(0\alpha\beta)}(\mathbf{x}/\varepsilon)$ the local stresses, corresponding to the global deformation $u_{\alpha,\beta x}(\mathbf{X}) = \delta_{\alpha\beta}$ in the plane of the plate, and denote $\sigma_{ij}^{(1\alpha\beta)}(\mathbf{x}/\varepsilon) = a_{ijkl} n_{k,ly}^{(1\alpha\beta)}(\mathbf{x}/\varepsilon)$ local stresses corresponding to the global curvatures $w_{,\alpha x\beta x}(\mathbf{X}) = \delta_{\alpha\beta}$.

In the joint and its vicinities, these stresses are determined by solving the local problem (15), and in the main parts of the plates through local solutions of problem (15). The local problem (15) can be solved analytically [Kolpakov, 2004], and the local problem (14) numerically. By (4) and (13), the local stresses in the plate are

$$\sum_{\alpha,\beta=1,2} [\sigma_{ij}^{(0\alpha\beta)}(\mathbf{x}/\varepsilon) u_{\alpha,\beta x}(\mathbf{X}) + \sigma_{ij}^{(1\alpha\beta)}(\mathbf{x}/\varepsilon) w_{,\alpha x\beta x}(\mathbf{X})]. \qquad (29)$$

Formulas (13) and (29) allow us to obtain displacements and local stresses in the plate in the form of a linear combination of the functions of "fast" and "slow" variables. These functions are determined from different problems: three-dimensional local problems and the two-dimensional problem of the theory of plates.

The values $u_{\alpha,\beta x}(\mathbf{X})$ ($\alpha,\beta = 1,2$) form the basis of two-dimensional strains in the plane of the plate, and values $w_{,\alpha x\beta x}(\mathbf{X})$ ($\alpha,\beta = 1,2$) form the basis of the curvatures and torsion of the plate. Accordingly, the functions $\mathbf{N}^{(0\alpha\beta)}(\mathbf{x}/\varepsilon)$, $\mathbf{N}^{(1\alpha\beta)}(\mathbf{x}/\varepsilon)$ and the stresses $\sigma_{ij}^{(0\alpha\beta)}(\mathbf{x}/\varepsilon)$, $\sigma_{ij}^{(1\alpha\beta)}(\mathbf{x}/\varepsilon)$ ($\alpha,\beta = 1,2$) form the basis of local displacements and stresses.

By virtue of formulas (13) and (29), we provide a complete decomposition of the original problem. According to these formulas, in order to solve the original problem (1), it is necessary to solve local problems and to determine the basic functions $\mathbf{N}^{(0\alpha\beta)}(\mathbf{x}/\varepsilon)$, $\mathbf{N}^{(1\alpha\beta)}(\mathbf{x}/\varepsilon)$, $\sigma_{ij}^{(0\alpha\beta)}(\mathbf{x}/\varepsilon)$, $\sigma_{ij}^{(1\alpha\beta)}(\mathbf{x}/\varepsilon)$ ($\alpha,\beta = 1,2$). Solving the problem of the theory of plates, it is necessary to find the functions $u_{\alpha,\beta x}(\mathbf{X})$ and $w_{,\alpha x\beta x}(\mathbf{X})$. Then, it is possible to calculate the local stresses from (29).

Note that the functions $\mathbf{N}^{(0\alpha\beta)}(\mathbf{x}/\varepsilon)$, $\mathbf{N}^{(1\alpha\beta)}(\mathbf{x}/\varepsilon)$, $\sigma_{ij}^{(0\alpha\beta)}(\mathbf{x}/\varepsilon)$, $\sigma_{ij}^{(1\alpha\beta)}(\mathbf{x}/\varepsilon)$ ($\alpha,\beta = 1,2$) are determined only by the geometry and the material properties of joint elements, and they are independent of the forces applied to the joined plates "as a whole", i.e. they are only individual properties of the joint.

The functions $u_\alpha(\mathbf{X})$ ($\alpha = 1,2$) and $w(\mathbf{X})$ are found by solving the problem of the theory of plates, i.e. determined by stresses and boundary conditions "in a whole". Since the limit problem of the plate theory does not depend on the individual properties of the joint, the functions $u_\alpha(\mathbf{X})$ ($\alpha = 1,2$) and $w(\mathbf{X})$ are also independent of the individual properties of the joint.

In other words, the problems of the deformation of joints and the deformation of connected plates "in a whole" are independent and the resulting decomposition is indeed complete and subdivides the original problem into two independent problems.

Below, we consider some typical types of plate joints and present examples illustrating the application of the method developed for these cases. One feature of the method is to find the solutions of local problems. Typically, these solutions can only be obtained numerically. By using modern codes such as ANSYS, NASTRAN and similar, the numerical solution of these problems is a common engineering task in the sense that the local problems do not contain a small parameter and other features requiring a rigorous scientific analysis.

## 2 Joint of constant cross-section

If the geometry and the material characteristics of the plates and the joint do not depend on the variable $x_3$ as well as the mass forces and boundary conditions, the three-dimensional problems of elasticity are replaced by two-dimensional ones. This also applies to the local problems. As an example, consider two connected plates as shown in Fig. 2. This is the model of a two-sided weld joint.

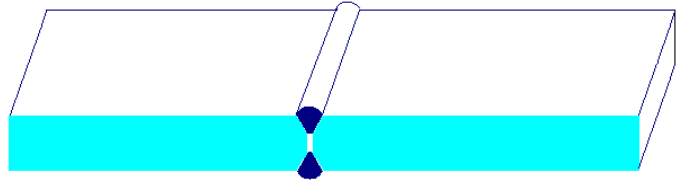

**Fig. 2.** Joint of constant cross-section between two homogeneous plates.

Fig. 3 shows the results of numerical solutions of the boundary value problem (14) (numerical solution of the problem carried out with the help of the program ANSYS [www.ansys.com, 2001]). The intensity of the stresses and the deformed shape of the bending plates in a direction perpendicular to the joint at different scales are shown. The upper diagram in Fig. 3a shows the loading scheme; in Fig. 3b one can see the deformed shape of the plate (view from the end). In Fig. 3c, we show the detailed structure of the intensity distribution of local stresses in the domain of the joint and in its vicinity. It can be seen that the perturbation of stresses is localized in the area, the length of which is of the order of the plate thickness. Note that in the present case, the joint itself occupies a narrow domain and the perturbation of the solution looks similar to a boundary layer. In the next section, we present an example of a joint occupying a relatively large domain.

In the computations, the following material characteristics have been used: for plate Young's modulus is $E_1$=200GPa, Poisson's ratio is $\nu_1$=0.2; for joint Young's modulus is $E_2$=230GPa, Poisson's ratio is $\nu_2$=0.3. Thickness of plate is 0.01m.

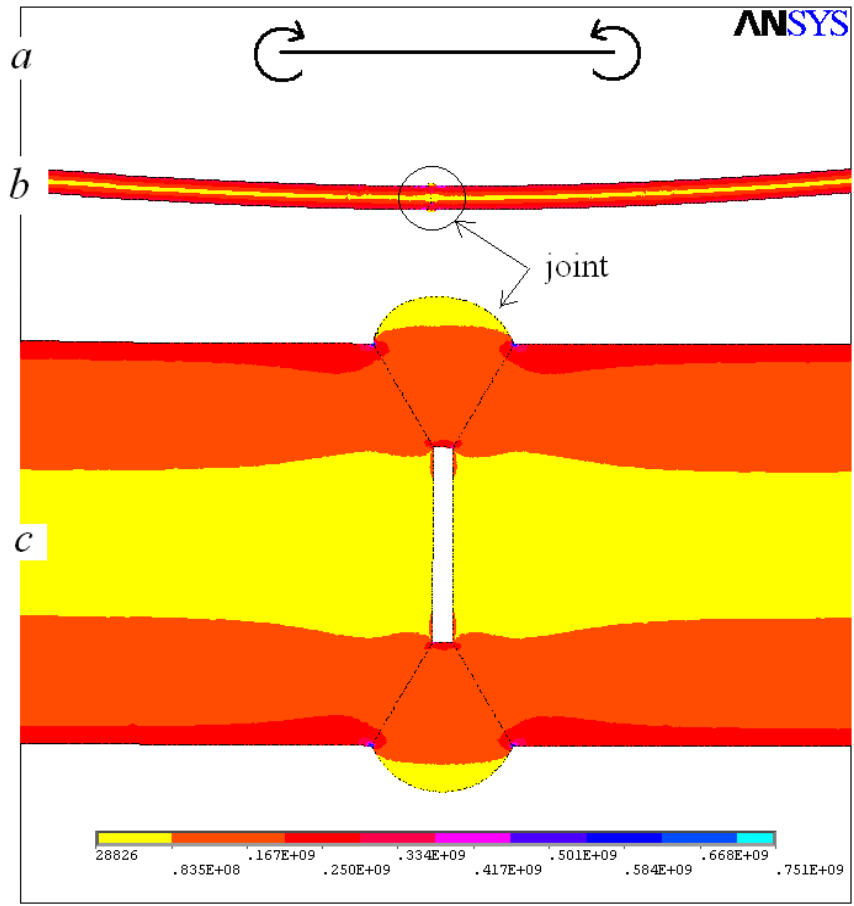

**Fig.3.** From top to bottom: loading scheme (a), the shape of the deformed plate (b), the local stresses intensity in the joint and its neighborhood (c).

## 3 A joint of periodic structure

In real structures joins often are periodic in the direction along the joint, see Fig.4, which shows a model of discrete welded joint.

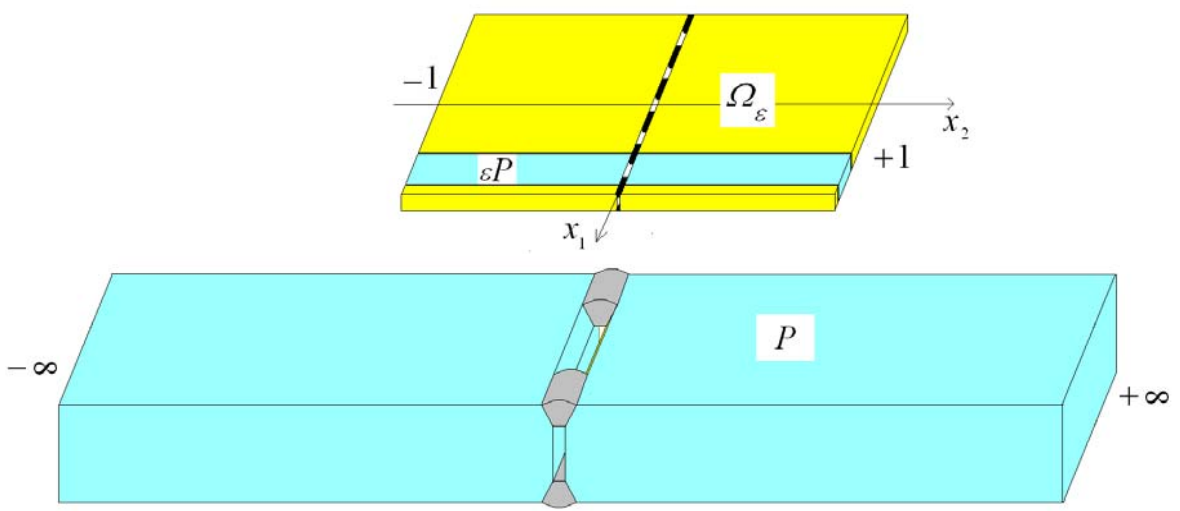

**Fig. 4.** Joint of the periodic structure (top) and the periodical cell (bottom).

The periodic structure of the joint causes the periodic nature of the stress-strain state of the whole structure. We denote the period of the structure through $\varepsilon T$.

In natural – "slow" variables $\mathbf{x}$, dimension of the periodicity cell is $\varepsilon P$, in the direction of $Ox_1$ - axis, along the joint, is $\varepsilon T$; in the direction of $Ox_2$ the axis cell extends the entire length of the plate (see Fig. 4). In "fast" variables $\mathbf{y}$, periodicity cell $P$ has wide $T$ and extends from $-\infty$ to $\infty$ in the direction $Oy_2$ (remind that $y_2 = x_2/\varepsilon$). In the case under consideration, we use representation (2) for the displacements with additional condition: $\mathbf{v}(\mathbf{y})$ is periodic in $y_1$ with period $T$. Correspondingly, the last condition in (14) (condition of the solution stabilization at infinity) must be carried out in the direction perpendicular to the joint, and in the direction parallel the joint solution must be periodic.

I.e., function $\mathbf{v}_l(\mathbf{y})$ is localized in a direction perpendicular to the joint and is periodic in a direction parallel to the joint. As a result, one obtains the following boundary value problem on the periodical cell $P$ ($\nu = 0,1$):

$$\begin{cases} \{a_{ijkl} N_{k,ly}^{\nu\alpha\beta}(\mathbf{y}) - a_{ij\alpha\beta} y_3^\nu\}_{,jy} = 0 \text{ in } P, \\ \{a_{ijkl} N_{k,ly}^{\nu\alpha\beta}(\mathbf{y}) - a_{ij\alpha\beta} y_3^\nu\} n_j = 0 \text{ on the free surface}, \\ \mathbf{N}^{(\nu\alpha\beta)}(\mathbf{y}) \to \mathbf{n}^{(\nu\alpha\beta)}(y_3) \text{ as } |y_2| \to \infty, \\ \mathbf{N}^{(\nu\alpha\beta)}(\mathbf{y}) \text{ periodic in } y_1 \text{ with period } T. \end{cases} \quad (30)$$

The problem (30) is a combination of the local problem (14) (stabilization of solutions to the variable $y_2$) and cell problem of the homogenization theory (periodicity to the variable $y_1$).

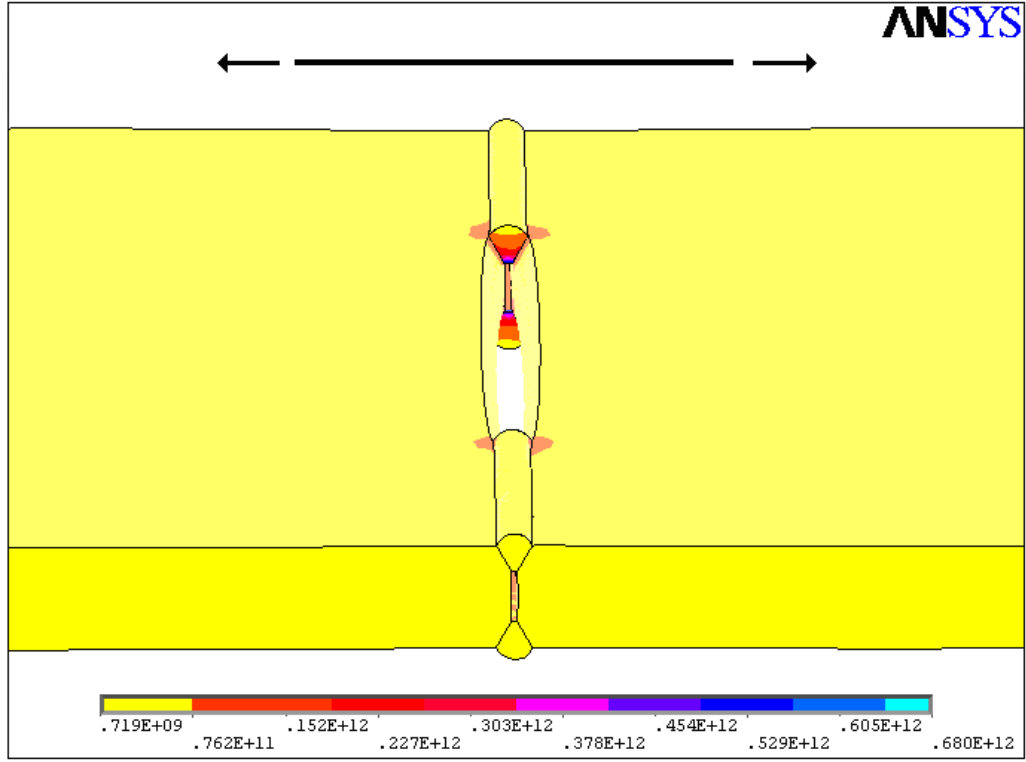

**Fig. 5.** Loading scheme (top) and the local stresses intensity in the joint (bottom).

Fig. 5, 6 shows the results of the numerical solutions of boundary value problem (30) for two plates with a joint of the periodic structure displayed in Fig. 4. Fig. 5 shows the diagram of loading, deformed shape and stress intensity for solution to cell problem corresponding to tension plate in its plane in a direction perpendicular to the joint. It is seen that the perturbation of stresses is localized near the joint. Fig. 6 shows the diagram of loading (Fig. 6a), bending form (Fig. 6b) and intensity of local stresses (Fig. 6c) for solution to the cell problem corresponding to bending of the plate. The area of localization of the stresses has the order of the size of the join.

In the computations, material characteristics were the following: for plate: Young's modulus is $E_1$ =200GPa, Poisson's ratio is $v_1$ =0.2; for joint Young's modulus is $E_2$ =230GPa, Poisson's ratio is $v_2$ =0.3. The thicknesses of the plates are 0.01 m.

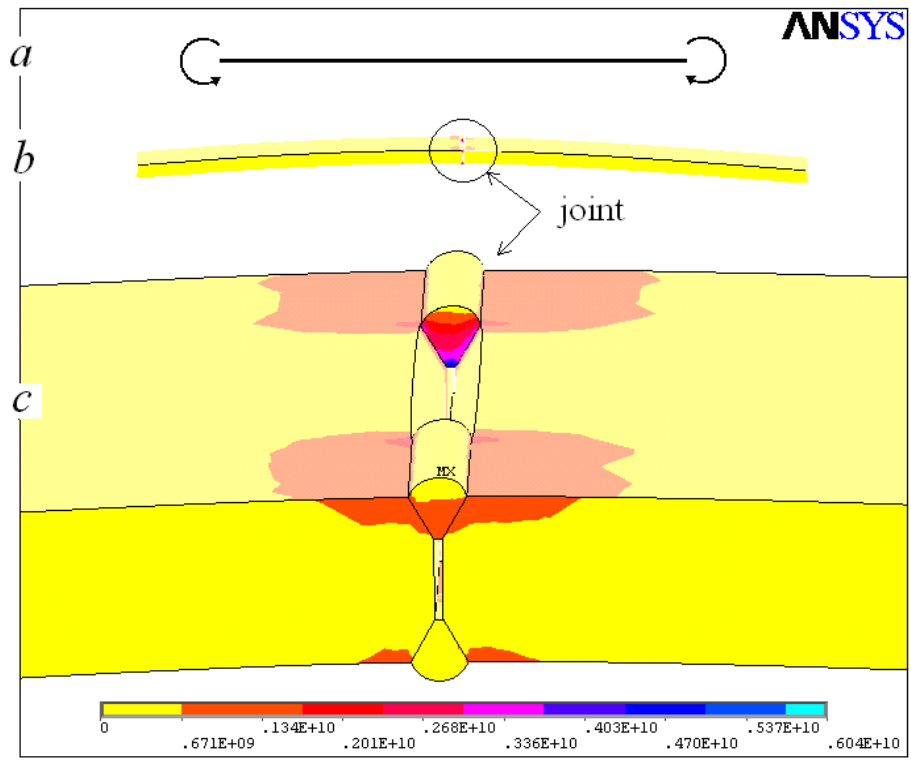

**Fig. 6**. Loading scheme (a), the deformed shape of the cell (b) and
the local stresses intensity in the joint (c).

## 6 Isolated (far located one from another) elements of joints

If the elements of joints (rivets, bolts, etc.) are located far enough from each other, they do not influence each other and the stress-strain state in each coupling element can be calculated independently of the others (one can see an analogy with dilute composites [Maxwell, 1873; Rayleigh, 1892]). At the same time, if the connectors are located close enough, they influence each other and the stress-strain state in each coupling element may be different from the stress-strain state calculated for an isolated element.

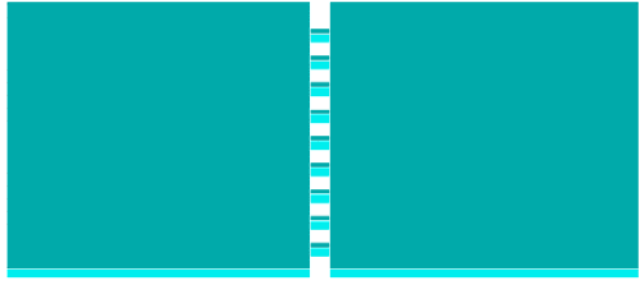

**Fig. 7.** Model of the system of joints.

The practical importance of accounting for the interaction of elements of joints or justification of the absence of interaction between the elements is obvious. The question comes down to what is "enough" close or "enough" far, respectively, in this problem. To answer the question, we consider a model problem–two identical uniform plates, connected by a system of joints, see Fig. 7.

Fig. 8 shows the stress intensity for the plate bended in the direction perpendicular to the joint (half of the plate is shown). In Fig. 8, the distance between the joints is five times the plate thickness. This is a limiting case.

Fig. 8 shows that the areas of stress perturbations are visibly not intersecting (it should be borne in mind that a discrete set of colors in the palette of ANSYS [www.ansys.com, 2001] slightly reduces the area of stress perturbations). In order to have a guaranteed safety margin, we increase the interval "five times the plate thicknesses" twice. Thus, the elements of the joint that are separated from each other by a distance greater than 10 times the plate thickness can be considered as isolated, do not interact with other elements of the joint. Joint elements spaced apart by a distance less than three thicknesses of the plate will tend to interact with each other and should together count as one integral joint element.

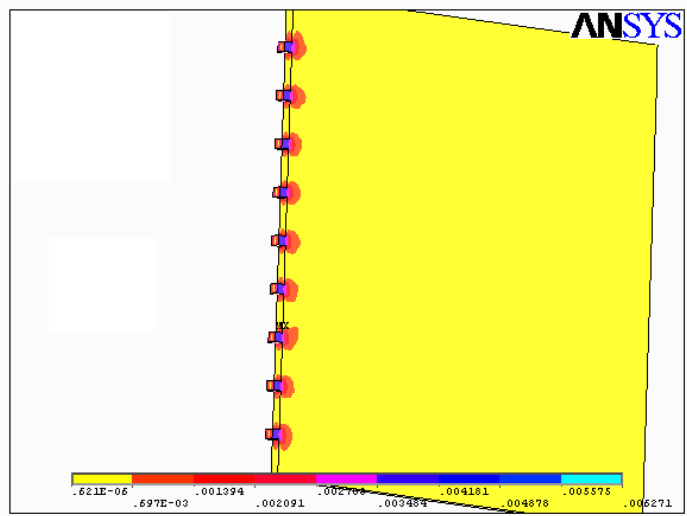

**Fig. 8.** The local stresses intensity in the system of connecting elements.

Next, we discuss the effect of "disappearance" of the joint in the limit problem, which has been already discussed theoretically in Section 1 (see formulas (21), (22) and the discussion on these formulas and the integrals (18), (19)). In Fig. 7, the main parts of the plates are the same. Since the connected plates are the same, all of the coefficients in (21) and (22) do not depend on **X** and, as the plates are homogeneous, can be consid-ered $C_{\alpha\beta\gamma\delta}(\mathbf{X}) = 0$. Then, the two-dimensional limit equilibrium equations (27) and (28) take the form

$$[-A_{\gamma\delta\alpha\beta}(\mathbf{X})u_{\alpha,\beta x}(\mathbf{X})]_{,\delta x} = f_\gamma(\mathbf{X}) \text{ in } D_1 \text{ and } D_2,$$

$$[A_{\alpha\beta\gamma\delta}(\mathbf{X})w_{,\alpha x\beta x}(\mathbf{X})]_{,\gamma x\beta x} = f_3(\mathbf{X}) \text{ in } D_1 \text{ and } D_2,$$

and the condition on the connecting line is: jump of values $[-D_{\gamma\delta\alpha\beta}w_{,\alpha x\beta x}(\mathbf{X})]n_\delta$, $[-D_{\gamma\delta\alpha\beta}w_{,\alpha x\beta x}(\mathbf{X})]_{,\delta x}n_\gamma$, and $[A_{\alpha\beta\gamma\delta}u_{,\alpha x\beta x}(\mathbf{X})]n_\delta$ on the line $L$ equal zero (Greek indices take values 1, 2). Domains $D_1$ and $D_2$ and line $L$ are displayed in Fig. 1. The differential equations and jump conditions above with the condition that $w \in C^1$ and $\mathbf{u} \in C$ means that the differential equations above are satisfied in the whole domain $D_1 \cup D_2$, see, e.g., [Fichera, 1972]. As a result, the limit problem consists of the problem of in-plane deformation of the plate and the problem of bending of the homogeneous plate. In both problems no transmission conditions arise on the perforated area between the plates.

The deflection of the plate is shown in Fig. 7, and the corresponding homogeneous plate will match up to the value of the order of the plate thickness (that would be the case and for reducing the three-dimensional homogeneous plate to its two-dimensional model). The stresses predicted by the two-dimensional model will be constant throughout the region, including the perforated area. In this case, the real stresses are shown in Fig. 7 – they are not constant.

On the connected line, both in the two-dimensional model and for the three-dimensional model shown in Fig. 7, the jump of the variables $M_{\alpha\beta}(\mathbf{X})n_\delta$ and $N_{\alpha\beta}(\mathbf{X})n_\delta$ is equal to zero (in the latter case this condition is satisfied for the mean values).

## 5 Connected plate of periodic structure

The local perturbation method, based on the idea of considering the physical and/or geometrical irregularities in a uniform external field, is well combined with the homogenization method, based on the same idea, but for the periodic system of irregularities rather than for an isolated physical heterogeneity. As a result, combination of the method of homogenization and the local perturbation method provides us with an effective method for the analysis of joined plates of periodic structure.

In the analysis of joined plates of periodic structure, the following representation for the displacements is used:

$$\mathbf{u}^\varepsilon(\mathbf{x}) = u_\alpha(\mathbf{X})\mathbf{e}_\alpha + w(\mathbf{X})\mathbf{e}_3 - w_{,\alpha x}(\mathbf{X})x_3\mathbf{e}_\alpha + \varepsilon\mathbf{V}(x_3/\varepsilon) + \varepsilon\mathbf{v}(\mathbf{x}/\varepsilon) = \quad (31)$$
$$= u_\alpha(\mathbf{X})\mathbf{e}_\alpha + w(\mathbf{X})\mathbf{e}_3 - \varepsilon w_{,\alpha x}(\mathbf{X})y_3\mathbf{e}_\alpha + \varepsilon\mathbf{V}_p(\mathbf{y}) + \varepsilon\mathbf{v}_l(\mathbf{y}).$$

In (31), the function $\mathbf{v}_l(\mathbf{y})$ is localized in the direction perpendicular to the direction of the joint, i.e., this is a local corrector. Function $\mathbf{V}_p(\mathbf{y})$ is periodic with respect to $y_1, y_2$ ($Oy_3$ axis is perpendicular to the plane of the plate) with the periodicity cell $P_l$ in the left plate and the periodicity cell $P_r$ in the right plate. This is a periodic corrector of the homogenization theory in the version for plates, see [Caillerie, 1984; Kohn and Vogelius, 1984; Kalamkarov and Kolpakov, 1997; Kolpakov, 2004].

As an example, we present the solution of the model problem of joined elastic perforated plates of periodic structure. The left plate has a square periodicity cell $P_3$ of size $3\text{mm} \times 3\text{mm}$ with a hole of diameter 2 (Fig. 9). The right-hand plate has a square periodicity cell $P_7$ of size $7\text{mm} \times 7\text{mm}$ with a hole of diameter 6. The periodic structure of the plates generates a periodic structure of the construction as a whole. The periodic cell $P$ of the construction has a size of 21 in the direction of the rib (21 is the least common multiple of the numbers 3 and 7) and extends in a direction from $-\infty$ to $\infty$ in the direction of the axis $Oy_2$. The thickness of the plate is 1mm.

Fig. 9 shows the results of the numerical calculation of the stresses in the plate. Away from the joint, the stresses have a periodic structure predicted by the homogenization theory. The stress-strain state in the joint region is very different from the stress-strain state in the plates (see Fig. 9, right). In this problem, the boundary-layer technique would probably not be effective.

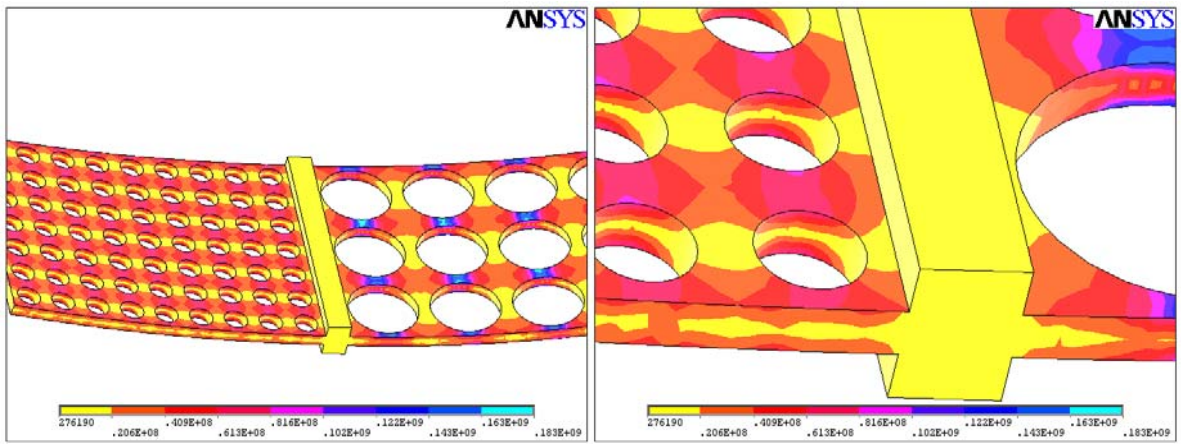

**Fig. 9.** Deformed joined plates of periodic structure and the local stresses intensity (left), zoomed (right).

In the problem under consideration, the condition of the decay of the solution away from the joint in the local problem, corresponding to the function $\mathbf{v}_l(\mathbf{y})$, can be replaced by the condition of the local solutions tending to the solution of the cell problem of the homogenization theory in the version for plates for a local problem, which corresponds to $\mathbf{V}_p(\mathbf{y}) + \mathbf{v}_l(\mathbf{y})$.

The first two equations in (14) are the same for all the local problems of the method of local perturbations (and homogenization theory too). The difference is in the third condition from (14) – condition to decay or stabilization of the solution away from the joint.

## 6 Conclusions

The main theoretical result of this paper consists in a justification of the possibility of *complete* decomposition of the problem of the deformation of joined plates: the separation of the original three-dimensional problem for the limiting two-dimensional problem of the theory of plates and the local three-dimensional problems. With regard to the partial decomposition of the problem connected to the thin-walled elements, see [Panasenko, 2005, 2007]. The above-mentioned two-dimensional problem is the leading term of the asymptotic and consists of two-dimensional equations of the theory of plates, in which the conditions of the contact line exist, but do not take into account the individual properties of the joint (either geometric or material characteristics of the elements of the connection).

Interaction conditions for the two-dimensional problem depend on the characteristics of the "main" parts of the plates (parts of the plates placed far away from the joint). The three-dimensional stress-strain state in the joint and in the vicinity of the joint is determined from the local problem solutions corresponding to this joint using the solution of the two-dimensional problem as parameter.

The local problem takes into account the individual properties of the joint and can be treated as a problem of determining the stress-strain state of an isolated connecting assembly in the global field of homogeneous strain, where a homogeneous deformations determined by solving the two-dimensional problem.

Note that if a structure contains many identical joints, than the solution of a local problem for one joint is sufficient for the calculation of the stress-strain state in each of the joints.

There raises a question important for practice: under what conditions can joints be considered as not mutually influencing each other (otherwise they should be considered as elements of a complex multi-element joint and analyzed together)? Based on the solutions of the model problem, we conclude that joints separated by a distance of more than 10 times of the plate thickness can be considered as separated. In contrast, joints which are located less than 3 times of the plate thickness from one another should be combined into one connector assembly.

The local problems (i.e., the problems of deformation of singular joints) are formally defined in infinite regions, but due to the effect of localization [Kolpakov and Kolpakov, 2010] (which for elasticity problems in thin-walled structural elements is a manifestation of the effect of de Saint-Venant), the solution is sufficient to perform in areas of a size 10 to 20 times of the thickness of the connected plates. The three-dimensional problem of the theory of elasticity in areas of these dimensions can be solved numerically for very complex geometries and an arbitrary distribution of the elastic characteristics by using standard codes such as ANSYS, NASTRAN, etc. This makes the method proposed in this work a practical method for the analysis of real engineering structures.

The proposed method is applicable for calculating homogeneous connected plates as well as connected plates of complex periodic structure.

The local perturbation method can be applied to the analysis of various isolated inhomogeneities (joints are special cases of inhomogeneities) in thin-walled structures of different types.

**Acknowledgments** The research was supported by Deutsche Forschungsgemeinschaft (DFG), grant no. WE 736/30-1 (for A.G. Kolpakov). A.G. Kolpakov thanks for the hospitality at the Institute of General Mechanics (IAM), RWTH Aachen University.

### Appendix. Solution to the problem (15)

It is simple to solve problem (15), actually more simple than the similar periodicity cell problem in the homogenization theory (it concerns problem (15) with index $v = 0$ (do not confuse this index with Poisson's ratio)), there is no analogue to problem (15) with index $v = 1$ in the homogenization theory of solids [Sanchez-Palencia, 1980; Jikov, 1994; Bensoussan et al., 1978]). This is due to the boundary condition in (15). At the same time, the solution of problem (15) is nontrivial even for a uniform plate made of homogeneous material (solution of the similar periodic cell problem in the homogenization theory is zero if the material is homogeneous).

From the differential equation in (15), we have

$$a_{i3k3}\frac{dn_k^{(v\alpha\beta)}}{dy_3}(y_3) - a_{i3\alpha\beta}y_3^v = C_i, \qquad (A1)$$

where $C_i$ ($i, k = 1, 2, 3$) are constants. From the boundary condition in (15), we find that $C_i = 0$. As a result, equation (A1) takes the form

$$a_{i3k3}\frac{dn_k^{(v\alpha\beta)}}{dy_3}(y_3) - a_{i3\alpha\beta}y_3^v = 0. \qquad (A2)$$

Solving (A2) with respect to $\dfrac{dn_k^{(\nu\alpha\beta)}}{dy_3}$, we have

$$\frac{dn_k^{(\nu\alpha\beta)}}{dy_3}(y_3) = a_{k3i3}^{-1}\, a_{i3\alpha\beta}\, y_3^{\nu}, \qquad (A3)$$

where "$-1$" indicates the inverse of the $3\times 3$ matrix $a_{i3k3}$. From equation (A3), the function $n_k^{(\nu\alpha\beta)}$ can be computed up to a constant. Note, that the derivative $\dfrac{dn_k^{(\nu\alpha\beta)}}{dy_3}$ and not the function $n_k^{(\nu\alpha\beta)}$ is used in the analysis of the plate.

From (A3), it is seen that both $\dfrac{dn_k^{(0\alpha\beta)}}{dy_3}$ and $\dfrac{dn_k^{(1\alpha\beta)}}{dy_3}$ are not zero. This reflects the fact that the plate has nonzero deformations in the direction normal to the plane of the plate both under the deformation in its plane and under bending.